\documentclass[pra,aps,10pt,superscriptaddress,twocolumn,floatfix]{revtex4-1}
\usepackage{graphicx,color}
\usepackage[nice]{nicefrac}							
\usepackage{amsmath,amssymb,bm}
\usepackage{ulem}

\usepackage[plainpages=false,pdfpagelabels,colorlinks=true,linkcolor=red,urlcolor=blue,citecolor=blue,pdftitle={},pdfauthor={},pdfdisplaydoctitle=true,pdfduplex=DuplexFlipLongEdge]{hyperref}

\definecolor{darkred}{rgb}{0.90,0.2,0.2}
\definecolor{darkgreen}{rgb}{0,0.60,.2}
\definecolor{darkblue}{rgb}{0.1,0.3,1}
\definecolor{grey}{cmyk}{0,0,0,0.25}
\definecolor{orange}{cmyk}{0,0.6,0.8,0}

\begin{document}

\title{Phenomenology of spectral functions in disordered spin chains at infinite temperature}

\author{Lev Vidmar}
\affiliation{Department of Theoretical Physics, J. Stefan Institute, SI-1000 Ljubljana, Slovenia}
\affiliation{Department of Physics, Faculty of Mathematics and Physics, University of Ljubljana, SI-1000 Ljubljana, Slovenia}
\author{Bartosz Krajewski}
\affiliation{Department of Theoretical Physics, Faculty of Fundamental Problems of Technology, Wroc\l aw University of Science and Technology, 50-370 Wroc\l aw, Poland}
\author{Janez Bon\v ca}
\affiliation{Department of Physics, Faculty of Mathematics and Physics, University of Ljubljana, SI-1000 Ljubljana, Slovenia}
\affiliation{Department of Theoretical Physics, J. Stefan Institute, SI-1000 Ljubljana, Slovenia}
\author{Marcin Mierzejewski}
\affiliation{Department of Theoretical Physics, Faculty of Fundamental Problems of Technology, Wroc\l aw University of Science and Technology, 50-370 Wroc\l aw, Poland}

\begin{abstract}
Studies of disordered spin chains have recently experienced a renewed interest, inspired by the question to which extent the exact numerical calculations comply with the existence of a many-body localization phase transition.
For the paradigmatic random field Heisenberg spin chains, many intriguing features were observed when the disorder is considerable compared to the spin interaction strength.
Here, we introduce a phenomenological theory that may explain some of those features.
The theory is based on the proximity to the noninteracting limit, in which the system is an Anderson insulator.
Taking the spin imbalance as an exemplary observable,
we demonstrate that the proximity to the local integrals of motion of the Anderson insulator determines the dynamics of the observable at infinite temperature.
In finite interacting systems our theory quantitatively describes its integrated spectral function for a wide range of disorders.
\end{abstract}

\maketitle

{\it Introduction.}
A considerable effort has been devoted to understanding the emergence of ergodicity in physically relevant quantum many-body systems.
Important cornerstones are provided by the random matrix theory (RMT) and the eigenstate thermalization hypothesis (ETH)~\cite{deutsch_91, srednicki_94, rigol_dunjko_08, dalessio_kafri_16, mori_ikeda_18, deutsch_18}.
Even though a rigorous proof of the ETH is still missing, several exact numerical studies confirmed its validity with remarkable accuracy, at least for specific parameter regimes of some physical Hamiltonians~\cite{dalessio_kafri_16, santos2010, Beugeling2014, Steinigeweg2014, Kim_strong2014, mondaini_rigol_17, jansen_stolpp_19, leblond_mallayya_19, mierzejewski_vidmar_20, brenes_leblond_20, richter_dymarsky_20, schoenle_jansen_21, brenes_pappalardi_21}.
The clearest numerical results have been obtained for the regimes where all model parameters are quantitatively similar and the numerical artifacts are strongly suppressed.
Much less understood are properties of many-body systems in which some physical processes (e.g., interaction or quenched disorder) are dominant over all other processes.
Exciting open questions concern the possibility of ergodicity breaking phase transitions and a generalization of the Kolmogorov-Arnold-Moser theorem~\cite{kolmogorov_54, caux_mossel_11, brandino_caux_15}.
In strongly disordered systems, this type of ergodicity breaking phase transition is referred to as the many-body localization transition~\cite{basko_aleiner_06, gornyi_mirlin_05, pal_huse_10, Rahul15, altman_vosk_15, alet_laflorencie_18, abanin_altman_19}.

A recent study~\cite{suntajs_bonca_20a} argued that the identification of ergodicity in numerical results may strongly depend on the value of the Thouless time $t_{\rm Th}$ relative to the Heisenberg time $t_{\rm H}$~\footnote{
The Thouless time $t_{\rm Th}$ may be seen as the longest physically relevant relaxation time, and the Heisenberg time $t_{\rm H}$ is proportional to the inverse level spacing.
}.
A system is interpreted as ergodic if $t_{\rm Th} \ll t_{\rm H} $, while in the opposite regime $t_{\rm Th} \gtrsim t_{\rm H}$ the interpretation of finite-size results appears to be less conclusive.
For a quantitative illustration, let us consider the random field Heisenberg chain with $L$ sites,
\begin{equation}
\hat H= J \sum_i (\hat S^x_i \hat S^x_{i+1} + \hat S^y_i \hat S^y_{i+1}+ \Delta \hat S^z_i \hat S^z_{i+1})  + \sum_i h_i \hat S_i^z, \label{hamh}
\end{equation}
where $\hat S_i^\alpha$ ($\alpha=x,y,z$) are standard spin-1/2 operators and the local fields $h_i$ (in units of $J\equiv 1$) are independent and identically distributed random variables drawn from the box distribution, $h_i \in  [-W,W]$.
It was shown~\cite{suntajs_bonca_20a} that in finite systems ($L \lesssim 20$) at $\Delta = 1$, the criterion $t_{\rm Th} \sim t_{\rm H}$ is satisfied around $W = W^* \approx 2$.
Considering the behavior of the system~(\ref{hamh}) with increasing disorder strength $W$, this point can therefore be interpreted as the onset of the ergodicity breakdown.
The latter is consistent with the level statistics and the eigenstate entanglement entropies departing from the RMT predictions~\cite{suntajs_bonca_20}, the fidelity susceptibility being maximal~\cite{sels2020}, the distribution of observable matrix elements being anomalous~\cite{panda_scardicchio_20, corps_molina_21}, the opening of the Schmidt gap~\cite{gray_bose_18} and the gap in the spectrum of the eigenstate one-body density matrix~\cite{bera_schomerus_15}, and the correlation-hole time in the survival probability reaching $t_{\rm H}$~\cite{schiulaz_torresherrera_19}.

Despite those developments, the fate of the ergodicity breaking point in the thermodynamic limit remains an extensively debated topic~\cite{suntajs_bonca_20a, suntajs_bonca_20, panda_scardicchio_20, sierant_delande_20, sierant_lewenstein_20, sels2020, abanin_bardarson_21}.
Moreover, previous studies reported other fascinating phenomena such as subdiffusive transport~\cite{barlev_cohen_15, agarwal_gopalakrishnan_15, luitz_laflorencie_16, khait_gazit_16, znidaric_scardicchio_16, luitz_barlev_17, bera_detomasi_17} and an approximate $1/\omega$ scaling of the spin density spectral function~\cite{mierzejewski2016, serbyn2017, sels2020}.
These observations call for a universal description within a simple theory that should provide quantitative predictions at all disorder strengths.

In this Letter we introduce a phenomenological theory that may achieve some of those goals.
We develop the theory on the premise that the noninteracting point at $\Delta = 0$, which is Anderson localized for any disorder in the thermodynamic limit~\cite{anderson_58, Mott1961}, determines specific properties of disordered spin chains also at $\Delta \ne 0$.
The key ingredient of the theory is the proximity to the local integrals of motion of the Anderson insulator (shortly, Anderson LIOMs).
In particular, we allow the Anderson LIOMs to acquire finite relaxation times due to interactions, i.e., they may become delocalized.
The theory provides an analytical description of the frequency dependence of the spectral function, it exhibits a remarkable agreement with numerical results for a wide range of disorders, and it suggests that at least a fraction of Anderson LIOMs are delocalized.
Specifically, for the spin imbalance observable, we explain rich phenomenology of the spectral function, which ranges from the anomalous $\approx 1/\omega$ behavior at moderate disorders to more complicated functional forms at strong disorder.

{\it Spectral function.}
The central quantity in our studies is the spectral function $S(\omega)$ of an observable $\hat A$, which is the Fourier transform of its autocorrelation function, 
\begin{equation}
S(\omega)=\frac{1}{2 \pi} \int_{-\infty}^{\infty} {\rm d} t \; e^{i \omega t -|t| 0^+} \langle e^{i\hat Ht} \hat A e^{-i\hat Ht} \hat A \rangle\;,
\label{iomega}
\end{equation}
where $\langle \cdots \rangle = {\rm Tr}\{\cdots\}/{\cal D}$ denotes the ensemble average over all eigenstates and ${\cal D}$ is the dimension of the Hilbert space.
Our numerical calculations are carried out for its integral
\begin{eqnarray}
I(\omega)=\int_{-\omega}^{\omega} {\rm d} \omega'  S(\omega')= \frac{1}{\cal D} \sum_{m,n=1}^{\cal D} \theta \left( \omega-|E_m-E_n| \right) A^2_{mn}\;,  \nonumber \\
\label{itomega}
\end{eqnarray} 
where $E_n$ are the energy levels and $A_{mn} \equiv \langle m | \hat A | n \rangle$ are matrix elements of $\hat A$ in the eigenstate basis,  $\hat H | n \rangle=E_n   | n \rangle$, $\theta$ is the Heaviside step function, and we set $\hbar \equiv 1$. 
We study observables that are traceless,
$\langle \hat A \rangle = 0$,
and normalized, 
$||\hat A||^2= \langle \hat A \hat A \rangle=1$~\cite{mierzejewski_vidmar_20}.
As a consequence, the high-frequency limit of $I(\omega)$ equals
$\lim_{\omega \to \infty }I(\omega)=  \frac{1}{\cal D} \sum_{m,n} A^2_{mn}=  \langle \hat A \hat A \rangle = 1$.

The integrated spectral function $I(\omega)$ filters out fast fluctuations and thereby allows for a robust analysis of the dynamics encoded in $I(\omega)$ even for a single realization of disorder.
A particular observable that we study is the spin imbalance,
$\hat A=\frac{2}{\sqrt{L}}\sum_i (-1)^i \hat S^z_i$.
This observable has been measured experimentally~\cite{schreiber15, lueschen_bordia_17}, it is a self-averaging quantity in macroscopic systems, and it has nonvanishing projections 
on multiple Anderson LIOMs.
In the language of~\cite{pandey_claeys_20}, this observable is integrability preserving in the noninteracting limit $\Delta = 0$.

\begin{figure}
\centering
\includegraphics[width=\columnwidth]{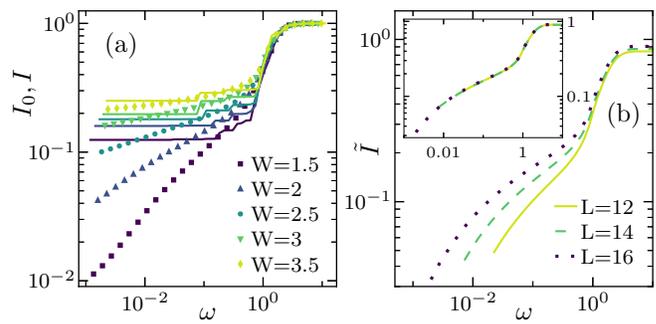}
\caption{
(a) Integrated spectral functions $I(\omega)$ [$\Delta=1$, symbols] and $I_0(\omega)$ [$\Delta=0$, lines] at $L=16$.
Results are shown for a single disorder realization and various values of $W$, such that the ratio $h_i/W$ in Eq.~(\ref{hamh}) is independent of $W$.
(b) 
Regular part $\tilde{I}(\omega)$, averaged over $10^3$ realizations of the disorder at $W=2$.
The results for $L=12$ and $14$ in the inset are shifted upwards by a constant to overlap with the data for $L=16$.
We set $J \equiv 1$ in all figures, and consider periodic boundary conditions in~(\ref{hamh}).
}
\label{fig1}
\end{figure}

{\it Comparison with the noninteracting limit.}
Figure~\ref{fig1}(a) shows $I(\omega)$ for a single realization of disorder at $\Delta=1$
(examples for other realizations are shown in~\cite{suppmat}). 
Results are compared to the noninteracting system, $I_0 (\omega)$ at $\Delta=0$.
For $\omega > J $ the results are qualitatively very similar, while important differences emerge in the low-frequency regime $\omega \ll J$, which is the main interest of this work.

The spectral weight of the Anderson insulator in the low-$\omega$ regime is strongly suppressed, which is manifested as $I_0 (\omega \ll  J) \simeq$ const.
This can be interpreted as the accumulation of the spectral weight of the observable in the stiffness $D_0=\lim_{\omega \to 0^+}I_0(\omega)$, and hence the spectral function can be approximated as $S_0 (\omega \ll J) \simeq D_0 \delta(\omega)$.
In contrast, the low-$\omega$ spectral weight of the interacting system may be considerable since $I (\omega \ll  J) \neq {\rm const}$.
This property gives rise to the {\it anomalous dynamics} of the imbalance for $\Delta \ne 0$ and $\omega \ll J$~\cite{znidaric_scardicchio_16, agarwal16, mierzejewski2016, luitz2016prl, gopal17, serbyn2017, prelovsek217, chanda2020, sels2020, prelovsek2021}, and is the main focus of this Letter.

As an important detail relevant for subsequent analysis, we note that the stiffness $D_0$ of an {\it arbitrary} observable $\hat A$ in the Anderson insulator ($\Delta=0$) originates from its projections on the Anderson LIOMs $\{\hat Q_{\alpha}\}$.
Therefore, the spectral function for $\omega \ll J $ can be written as
\begin{eqnarray}
S_{{\rm M},0}(\omega) = \sum_{\alpha} D_{\alpha} \delta( \omega)\,, \quad
 D_{\alpha} =\frac{\langle \hat A \hat Q_{\alpha} \rangle^2}{\langle \hat Q_{\alpha}   \hat Q_{\alpha} \rangle}\;,
 \label{mazur}
\end{eqnarray}
where $D_0 = \sum_{\alpha} D_{\alpha}$.
The latter relation follows from the Mazur bound~\cite{mierzejewski_vidmar_20}, and we consider the Anderson insulator as an integrable model containing orthogonal Anderson LIOMs  
$\langle \hat Q_{\alpha} \hat Q_{\alpha'} \rangle \propto \delta_{\alpha,\alpha'}$
(see~\cite{suppmat} for details about the Anderson LIOMs).
Since the projections $D_{\alpha}$ are defined in Eq.~(\ref{mazur}) by the average over the entire Hilbert space, we do not study the energy-resolved spectral functions, but instead we focus on the infinite temperature at which the average energy $(E_m+E_n)/2$ of pairs of eigenstates $| m \rangle, | n \rangle$ in Eq.~(\ref{itomega}) is arbitrary.

{\it Low-frequency regime.}
In what follows we focus on the interacting systems ($\Delta = 1$), and we disentangle the effect of accumulation of spectral weight in the stiffness from the low-$\omega$ spectral weight.
To this end, we study the {\it regular} part of the integrated spectral function, defined as $\tilde{I}(\omega) = I(\omega) 
-\frac{1}{\cal D} \sum_{n=1}^{\cal D}  A^2_{nn}$.
An example of the disorder averaged $\tilde{I}(\omega)$ at $W=2$ and different system sizes $L$ is shown in Fig. \ref{fig1}(b).
It is remarkable that a simple upward  shift of the curves for $L=12$ and $14$ results in an accurate overlap with the data for $L=16$.
This is observed at $W=2$ in the inset of Fig.~\ref{fig1}(b), and other values of the disorder in~\cite{suppmat}.
This suggests that the finite-size effects in the low-$\omega$ regime are small (apart from the $L$-dependent vertical shift), and calls for a simple theory to describe the observable spectral function.

An interesting remark can be made about the overlap of integrated spectral functions such as the one in the inset of Fig.~\ref{fig1}(b).
It indicates that a fraction of the spectral weight from the diagonal matrix elements at $\delta(\omega)$ is transferred to nonzero frequencies with increasing $L$.
This may be interpreted as the trend towards restoring the ergodicity in the thermodynamic limit.
Several works have recently explored possibilities for restoring the ergodicity at large disorders when the thermodynamic limit is approached~\cite{suntajs_bonca_20a, suntajs_bonca_20, kieferemmanouilidis_unanyan_20, kieferemmanouilidis_unanyan_21, sels2020, leblond2020}.
Nevertheless, our main focus here is to provide quantitative predictions for properties in {\it finite} systems.

{\it Proximity to Anderson insulator.}
We now construct a phenomenological theory that may quantitatively describe the observable spectral functions in finite systems.
Our approach is based on the proximity to the Anderson insulator whose conserved quantities are denoted as Anderson LIOMs.
Anderson LIOMs considered here do not imply existence of $l$-bits in interacting systems~\cite{huse14, Serbyn2013, ros15, chandran15, imbrie_16, thomson_schiro_18, detomasi_pollmann_19, kelly_nandkishore_20}.
The key premise of the theory is the conjecture that upon interactions, at least a fraction of Anderson LIOMs $\{\hat Q_{\alpha} \}$
become delocalized, i.e., they cease to be conserved and $\langle\hat Q_{\alpha}(t) \hat Q_{\alpha} \rangle$
decays with a finite relaxation time $\tau_{\alpha}$.
This impacts the dynamics of finite systems by broadening the $\delta$-functions in Eq.~(\ref{mazur}).
We model this effect by the following regular part of the spectral function for interacting system
[cf.~Eq.~(\ref{mazur})],

\begin{eqnarray}
S_{{\rm M}}(\omega \ll J )&=&\sum_{\alpha=1}^{N}   D_{\alpha}  \frac{1}{\pi} \frac{\tau_{\alpha}}{(\omega \tau_{\alpha})^2+1}\;,
 \label{mazur1}
\end{eqnarray}
where the summation runs over $N$ Anderson LIOMs that have nonvanishing projections on $\hat A$ and are delocalized in the interacting system.
Note that the broadening in Eq.~(\ref{mazur1}) is described by the Lorentzian functions, which is a common approach in the literature.
Recently, the Lorentzian form of the spectral function [cf.~Eq.~(\ref{mazur1}) with $N=1$] was actually observed in numerical studies of several many-body systems close to integrable points~\cite{mierzejewski2015, schoenle_jansen_21, leblond2020}.
Nevertheless, we argue in~\cite{suppmat} that the main results of our study are independent of the particular functional form of the broadening function.

Important inputs to the theory are the values of the stiffnesses $\{D_\alpha\}$ and the relaxation times $\{\tau_\alpha\}$ of delocalized Anderson LIOMs in the Hamiltonian~(\ref{hamh}).
We calculated both quantities numerically at disorders $W=2$ and 3, see Sec.~S4 of~\cite{suppmat}.
The first insight is that, for the spin imbalance, many projections
$D_\alpha$ from Eq.~(\ref{mazur}) are nonzero, and hence one needs to consider $N \gg 1$ in Eq.~(\ref{mazur1}).
The second insight is that the projections $D_\alpha$ are very weakly correlated (or uncorrelated) with the relaxation times $\tau_\alpha$, and hence we replace $D_\alpha$ with its average value in Eq.~(\ref{mazur1}),
$D_{\alpha} \to 1/N \sum_{\alpha} D_{\alpha} =D_0/N $.
Finally, we calculated the distribution $f_\tau(\tau)$ of the relaxation times $\tau_\alpha$ of the autocorrelation functions $\langle\hat Q_{\alpha}(t) \hat Q_{\alpha} \rangle$ and found that the distribution $f_\tau(\tau)$ is extremely wide.
In particular, the distribution can be well approximated by a power-law dependence $f_{\tau}(\tau) \propto 1/\tau^\mu$ in an interval $\tau \in [\tau_{\rm min},\tau_{\rm max}]$, where the disorder strength only impacts the exponent $\mu$ and the boundaries $\tau_{\rm min}$ and $\tau_{\rm max}$.
Such a power-law distribution of relaxation times $\tau_\alpha$ is consistent with the distributions of $\tau_\alpha$ studied for the Anderson insulators coupled to regular bosons or hard-core bosons via the Fermi golden rule~\cite{mierzejewski2018_1, mierzejewski2019}.

Summarizing the above considerations, we replace the sum $N^{-1} \sum_{\alpha=1}^N$ in Eq.~(\ref{mazur1}) with the integral $\int_{\tau_{\rm min}}^{\tau_{\rm max}} {\rm d}\tau f_\tau(\tau)$, and obtain a phenomenological model to describe the low-frequency dynamics,
 \begin{eqnarray}
S_{\rm M}(\omega ) &=&  \frac{\bar D_0 }{\pi}  \int_{\tau_{\rm min}}^{\tau_{\rm max}} \frac{{\rm d} {\tau}}{\tau^{\mu-1}}  \frac{1}{(\omega \tau)^2+1}\;,
 \label{mazur2}
\end{eqnarray}
where $\bar D_0$ is a prefactor that determines the total spectral weight arising from the delocalized Anderson LIOMs.
In analogy to Eq.~(\ref{itomega}), we then define $\tilde I_{\rm M}(\omega)$ by the integral of $S_{\rm M}(\omega )$, see also~\cite{suppmat}.

\begin{figure}
\centering
\includegraphics[width=\columnwidth]{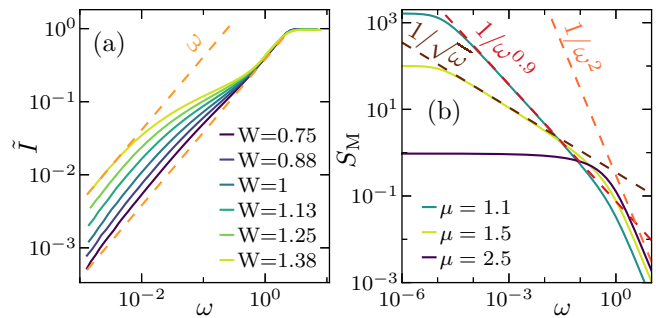}
\caption{ 
(a) Numerical results for the regular part of the integrated spectral function $\tilde I(\omega)$ at $L=16$ and weak disorder.
Results are averaged over $10^3$ realizations of disorder.
(b) Solid lines: $S_{\rm M}(\omega)$ from Eq. (\ref{mazur2}) at $\mu=1.1$, $1.5$ and $2.5$, using $\tau_{\rm min}=1$, $\tau_{\rm max}=10^5$ and $\bar D_0=1$.
Dashed lines are power-law guidelines, with functional forms $\propto 1/\omega^{2-\mu}$ for $\mu = 1.1, 1.5$, and $\propto 1/\omega^2$.
}
\label{fig2}
\end{figure} 

Before carrying out a quantitative comparison of our phenomenological model with the actual numerical data, we comment on some general properties of the spectral function described by Eq.~(\ref{mazur2}).
We first note that if $\omega \ll \tau_{\rm max}^{-1}$, then $S_{\rm M}(\omega) \propto {\rm const}$ and $\tilde I(\omega) \propto \omega$.
This property is usually associated with the diffusive character of the dynamics.
Emergence of such regime was detected in several studies of many-body systems that comply with the ETH~\cite{dalessio_kafri_16, dymarsky_18, brenes_leblond_20, brenes_goold_20, richter_dymarsky_20, leblond_rigol_20, schoenle_jansen_21, leblond2020}.
For the model under investigation, see Fig.~\ref{fig2}(a), we indeed observe $\tilde I(\omega) \propto \omega$ at $W \approx 1$.
In this regime of parameters, the phenomenological model~(\ref{mazur2}) can be simplified since $\tau_{\rm min}$ and $\tau_{\rm max}$ are of the same order and hence one may use a single relaxation time, $\tau_{\alpha} \to \tau$.
With increasing the disorder $W$, however, the linear regime in $\tilde I(\omega)$ shifts to lower $\omega$,  which is a consequence of a rapid increase of $\tau_{\rm max}$ with $W$.

\begin{figure}
\centering
\includegraphics[width=\columnwidth]{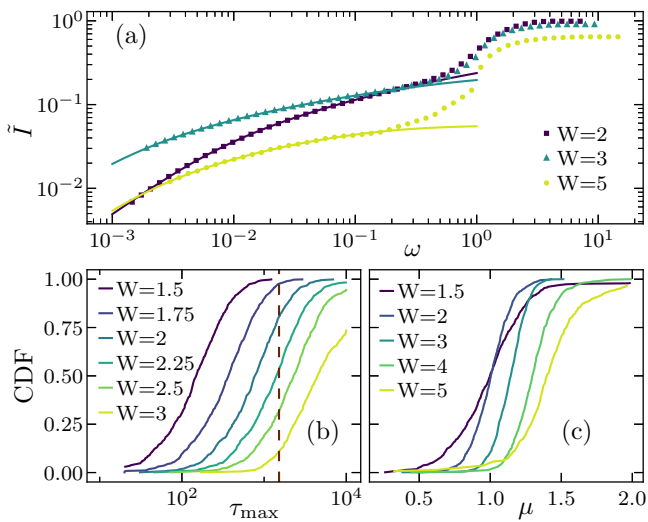}
\caption{
(a) Symbols: numerical results for $\tilde{I}(\omega)$ at $L=16$ and a single realization of the disorder $W$.
Lines: predictions by $\tilde{I}_M(\omega)$ for the low-frequency regime $\omega<0.2$. 
(b) and (c) The resulting cumulative distribution functions (CDF) of the fitting parameters $\tau_{\rm max}$ and $\mu$, respectively, for $10^3$ realizations of the disorder.
The vertical dashed line in (b) denotes the Heisenberg time $t_{\rm H}$ at $W=2$.
See~\cite{suppmat} for details.
}
\label{fig3}
\end{figure}

\begin{figure}
\centering
\includegraphics[width=\columnwidth]{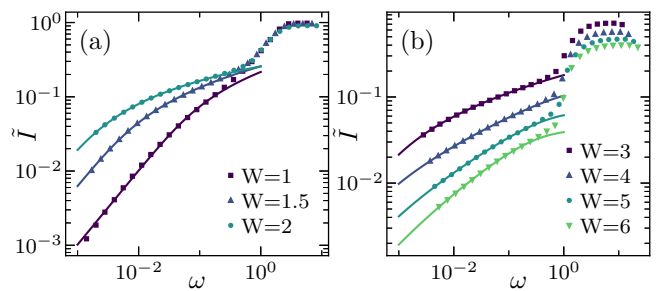}
\caption{
Symbols: numerical results for the disorder averages of $\tilde{I}(\omega)$ at $L=16$, using $10^3$ disorder realizations.
Lines: predictions by $\tilde{I}_M(\omega)$ for the low-frequency regime $\omega<0.2$.
Values of the disorder strengths are (a) $W \leq 2$ and (b) $W \geq 3$.
See~\cite{suppmat} for details.
}
\label{fig4}
\end{figure}

The main message of this Letter is that, for a wide range of disorder strengths, the low-frequency response may be governed by a broad distribution of the relaxation times $\{\tau_\alpha\}$, with $\tau_{\rm max}/\tau_{\rm min} \gg 1$ in Eq.~(\ref{mazur2}). This suggests that the frequency regime
$ \tau^{-1}_{\rm max} \ll \omega \ll   \tau^{-1}_{\rm min}$
may be very broad and hence relevant for the time regimes studied in numerical simulations and analog quantum simulators~\cite{schreiber15, lueschen_bordia_17}.
Particularly informative is the case $\mu=1$ in Eq.~(\ref{mazur2}), for which
 \begin{eqnarray}
S_{\rm M}(\omega )&=&  \frac{\bar D_0 }{\pi}\; \frac{{\rm arctan(\omega \tau_{\rm max} )-arctan(\omega \tau_{\rm min})} }{\omega}\,.
 \label{mazur3}
\end{eqnarray}
The functional form $\propto 1/\omega$ at $\mu=1$ is consistent with the anomalous dynamics and spectral functions reported in several previous studies~\cite{mierzejewski2016, serbyn2017, sels2020}.
More generally, $S_{\rm M}(\omega)$ at $\mu<2$ can roughly be approximated by $S_{\rm M}(\omega)\propto 1/\omega^\eta$ with $\eta \simeq 2-\mu$, see Fig.~\ref{fig2}(b) for $\mu = 1.1$ and $1.5$.
In~\cite{suppmat} we show that the $1/\omega^\eta$ dependence arises solely from the power-law distribution of relaxation times $\{\tau_\alpha\}$, and is not an artifact of the Lorentzian broadening used in Eq.~(\ref{mazur1}).
We note, however, that the functional forms predicted by Eq.~(\ref{mazur2}), as well as the numerical results in Figs.~\ref{fig3} and~\ref{fig4}, may also exhibit a fine structure beyond a simple power-law dependence.
In the opposite regime $\mu>2$, $S_{\rm M}(\omega)$ resembles a Fourier transform of a single Lorentzian, as shown in Fig. \ref{fig2}(b) for $\mu=2.5$.

{\it Numerical tests for spin imbalance.}
We now carry out a quantitative comparison between the numerical results for $\tilde{I}(\omega)$ [symbols in Figs.~\ref{fig3} and~\ref{fig4}] and the predictions $\tilde{I}_{\rm M}(\omega)$ from the phenomenological model in Eq.~(\ref{mazur2}) [lines in Figs.~\ref{fig3}(a) and~\ref{fig4}].
The fitting parameters of the latter are $\tau_{\rm min}$, $\tau_{\rm max}$ and $\mu$ that determine the distribution of relaxation times, and the prefactor $\bar D_0$.

Figure~\ref{fig3} considers the case where the free parameters of $\tilde{I}_{\rm M}(\omega)$ are fitted independently for every disorder realization.
An example of the outcome of such procedure is shown in Fig.~\ref{fig3}(a) for a single disorder realization, while examples for several other realizations are shown in~\cite{suppmat}.
Figures~\ref{fig3}(b) and~\ref{fig3}(c) then show the cumulative distribution of fitting parameters obtained by analyzing $10^3$ realizations of disorder.
There are two important quantitative results.
The first is that the distribution of $\tau_{\rm max}$ is broad and its median increases approximately exponentially with $W$, unless it reaches the Heisenberg time $t_{\rm H} = \omega_{\rm H}^{-1}$ at $W^* \approx 2$, see the vertical line in Fig.~\ref{fig3}(b).
(The Heisenberg energy $\omega_{\rm H}$ corresponds to the average level spacing in the middle of the spectrum, which at $L=16$ is $\omega_{\rm H}/J \approx 10^{-3}$~\cite{suntajs_bonca_20a}.)
The value $W^* \approx 2$ is consistent with the ergodicity breaking transition point in this model~\cite{suntajs_bonca_20}, occurring when the Thouless time $t_{\rm Th}$ in the spectral form factor approaches $t_{\rm H}$~\cite{suntajs_bonca_20a}.
When $\tau_{\rm max}$ exceeds $t_{\rm H}$, the mean of $\mu$ departs from $\mu=1$ towards higher values [see Fig.~\ref{fig3}(c)].
The second important result is that $\tau_{\rm min}$ remains well below $t_{\rm H}$ for all results reported here.
Otherwise, the dynamics would be frozen, $\tilde{I}(\omega)\simeq$ const, down to $\omega\sim \omega_{\rm H}$, which is clearly not the case in Figs.~\ref{fig3}(a) or \ref{fig4}(b).
The first result suggests that a fraction of Anderson LIOMs remains localized at $W > W^*$ upon adding the interactions.
Exploring the fate of those LIOMs for larger systems, i.e., when $t_{\rm H} \to \infty$, is beyond the scope of this work.
The second result suggests that at least some fraction of Anderson LIOMs is delocalized in the interacting system for all disorder values considered here.
In Fig.~\ref{fig4} we carry out an analogous analysis for the disorder averages of $\tilde I(\omega)$.
Also in this case, the phenomenological model from Eq.~(\ref{mazur2}) provides an extremely accurate description of the results.
A quantitative analysis of the fitting parameters $\tau_{\rm max}$ and $\mu$ is provided in~\cite{suppmat}.

{\it Conclusions}.
In this Letter we introduced a phenomenological theory that accurately describes the spectral properties of the spin imbalance in disordered  chains.
The theory is  based on the proximity to the Anderson insulator.
We assume that at least certain Anderson LIOMs acquire finite relaxation times as a consequence of interactions.
An important ingredient of the underlying phenomenological model is a broad distribution of relaxation times of Anderson LIOMs, which represents the origin of anomalous dynamics in finite systems.
Then in systems amenable to exact diagonalization there exist the disorder $W^*$ [$W^* \approx 2$ for the model in~(\ref{hamh})] above which the relaxation times $\{\tau_\alpha\}$ of a fraction of Anderson LIOMs are larger than the Heisenberg time $t_{\rm H}$.
As a result, the properties of finite systems at $W > W^*$ are governed by the coexistence of two types of LIOMs: those for which $\tau_\alpha > t_{\rm H}$ (they appear to be exactly conserved), and those for which $\tau_\alpha < t_{\rm H}$. 
The interplay between both types of LIOMs may give rise to unconventional properties of the system defined on a Fock space graph~\cite{deluca__scardicchio_13, Luitz2015, mace_alet_19, logan_welsh_19, roy_logan_20, detomasi_khaymovich_21}, which needs to be explored in more details in future work.

\acknowledgements
We acknowledge discussions with F. Heidrich-Meisner, D. Logan, A. Polkovnikov, P. Prelovšek, T. Prosen, M. Rigol, D. Sels and P. Sierant.
We acknowledge the support by the National Science Centre, Poland via project 2020/37/B/ST3/00020 (M.M.), the support by the Slovenian Research Agency (ARRS), Research Core Fundings Grants P1-0044 (L.V. and J.B.) and J1-1696 (L.V.), and the support from  the Center for Integrated Nanotechnologies, a U.S. Department of Energy, Office of Basic Energy Sciences user facility (J.B.). 

\bibliographystyle{biblev1}
\bibliography{references,references_ergtransition}
 

\newpage
\phantom{a}
\newpage
\setcounter{figure}{0}
\setcounter{equation}{0}

\renewcommand{\thetable}{S\arabic{table}}
\renewcommand{\thefigure}{S\arabic{figure}}
\renewcommand{\theequation}{S\arabic{equation}}
\renewcommand{\thepage}{S\arabic{page}}

\renewcommand{\thesection}{S\arabic{section}}

\onecolumngrid

\begin{center}

{\large \bf Supplemental Material:\\
Phenomenology of spectral functions in disordered spin chains at infinite temperature}\\

\vspace{0.3cm}

\setcounter{page}{1}

Lev Vidmar,$^{1,2}$
Bartosz Krajewski,$^{3}$
Janez Bon\v ca,$^{2,1}$ 
Marcin Mierzejewski$^{3}$  \\
\ \\
$^1${\it Department of Theoretical Physics, J. Stefan Institute, SI-1000 Ljubljana, Slovenia} \\
$^2${\it Department of Physics, Faculty of Mathematics and Physics, University of Ljubljana, SI-1000 Ljubljana, Slovenia} \\
$^3${\it Department of Theoretical Physics, Faculty of Fundamental Problems of Technology, \\ Wroc\l aw University of Science and Technology, 50-370 Wroc\l aw, Poland}\\

\end{center}

\vspace{0.6cm}

\twocolumngrid

\label{pagesupp}

\section{Details about Fig.~\ref{fig1}} \label{app1}

\begin{figure}[!b]
\centering
\includegraphics[width=\columnwidth]{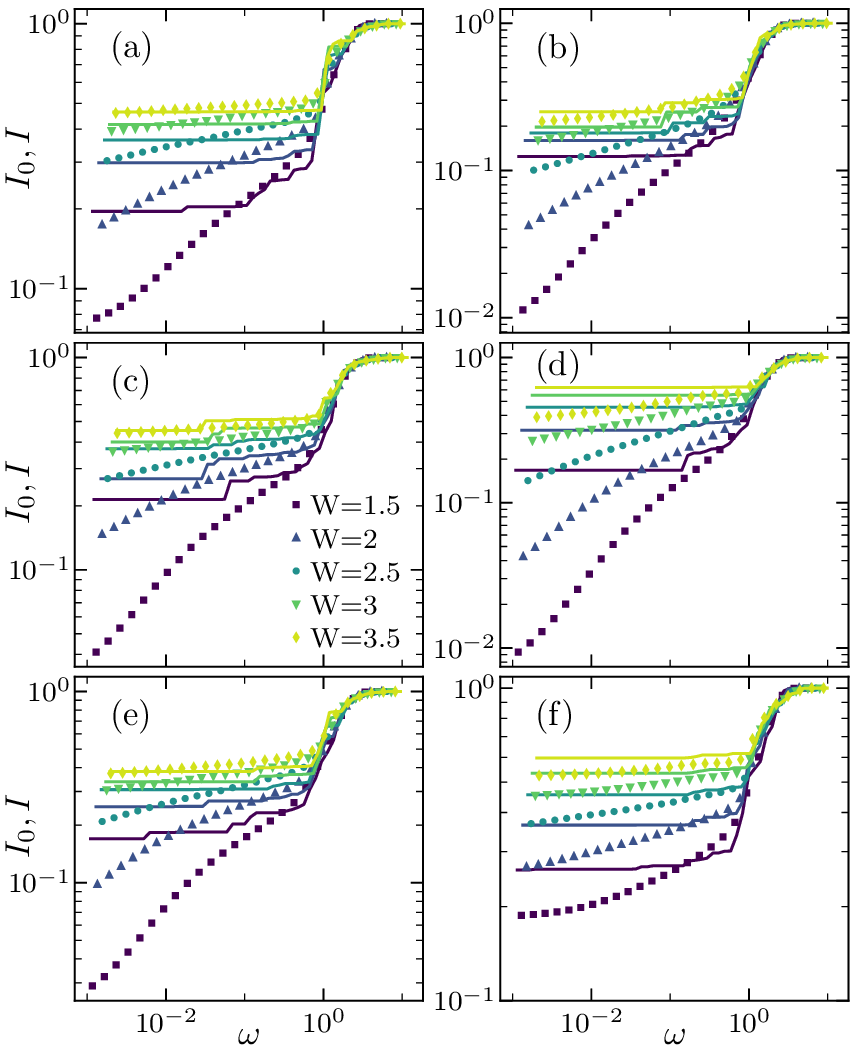}
\caption{
Integrated spectral function $I(\omega)$ at $\Delta=1$ (symbols) and its noninteracting counterpart $I_0(\omega)$ at $\Delta=0$ (lines), at $L=16$.
Each panel corresponds to a different realization of the disorder.
All results within a single panel are obtained for the same disorder realization, i.e., using identical values of the ratio $h_i/W$, for  $i=1,...,L$.
}
\label{figs1}
\end{figure} 

\begin{figure}[!b]
\centering
\includegraphics[width=\columnwidth]{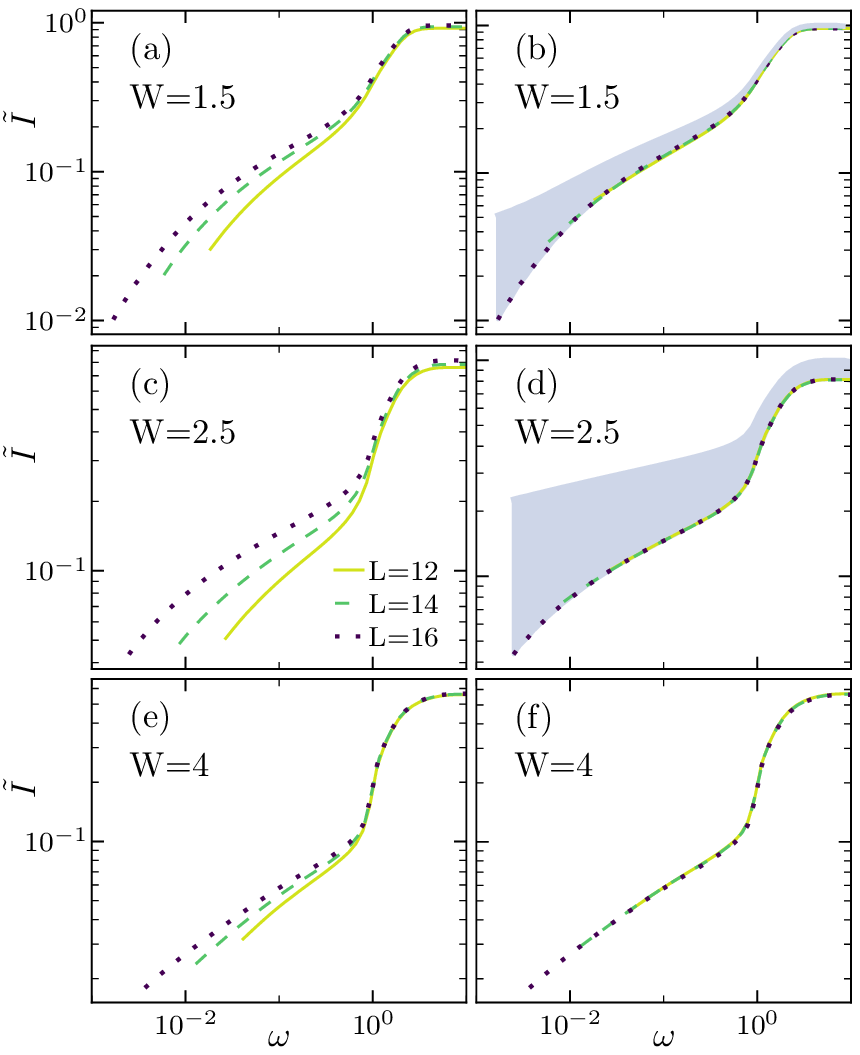}
\caption{
Regular part of the integrated spectral function $\tilde{I}(\omega)$, averaged over $10^3$ realizations of the disorder, for system sizes $L=12,14,16$ and distinct disorder amplitudes $W$.
(a), (c) and (e) show unscaled results, while in (b), (d) and (f) the results at $L=12$ and $14$ are shifted upwards by a constant (see Table~\ref{tabS1}) to overlap with the results at $L=16$.
Shaded areas in (b) and (d) show the estimates of $\tilde{I}(\omega)$ in the thermodynamic limit, see the text for details.} 
\label{figs2}
\end{figure} 

Figure~\ref{fig1}(a) of the main text shows the integrated spectral function $I(\omega)$, as well as its noninteracting counterpart $I_0(\omega)$ at $\Delta=0$, for a single realization of the disorder and different values of the disorder amplitude $W$.
In Fig.~\ref{figs1} we show those results for six other realizations of the disorder (we set $J \equiv 1$ in all figures).
All results share some common features: while the spectral weight at the noninteracting point $\Delta=0$ is strongly suppressed at $ \omega_{H}< \omega \ll J$ (i.e., $I_{0}\simeq$ const), it exhibits a nontrivial $\omega$-dependence in the interacting regime at $\Delta=1$.
For small values of $\omega$ close to the Heisenberg energy $\omega_{H}$, the integrated spectral functions in the interacting model are smaller than those at the noninteracting point.
On the other hand, the suppression of the spectral weight at nonzero but small energy $\omega \ll J$ at the noninteracting point supports Eq.~(\ref{mazur}) of the main text, which is the starting point for the phenomenological modeling of the spectral function in interacting systems. 

In Fig.~\ref{fig1}(b) of the main text we showed the regular part of the disorder averaged integrated spectral function $\tilde I(\omega)$ at $W=2$ and different system sizes $L=12,14,16$.
Results for the disorders $W=1.5$, $2.5$ and $4$ are shown in Figs.~\ref{figs2}(a),~\ref{figs2}(c) and~\ref{figs2}(e), respectively.
In all those cases, the unscaled results in the low-$\omega$ regime exhibit a robust $L$ dependence.
However, performing a vertical shift of the results at $L=12$ and $14$ by a constant,
$\tilde{I}(\omega) \to \tilde{I}(\omega) + \delta \tilde{I}$,
gives rise to an excellent overlap of the results.
The later is shown in the inset in Fig.~\ref{fig1}(b) [main text] and in Figs.~\ref{figs2}(b),~\ref{figs2}(d) and~\ref{figs2}(f), while
the values of $\delta \tilde{I}$ are given in the Table~\ref{tabS1}.

The overlap of shifted curves suggests that the spectral functions 
should have some universal properties. 
Unfortunately, the accessible system sizes do not allow for an unambiguous finite-size
scaling of the results shown in Table~\ref{tabS1}. 
However, one may still estimate $ \tilde{I}(\omega)$ in the thermodynamic
using the inequalities $ \tilde{I}(\omega) \le I(\omega) \le 1$.
Then for an arbitrarily large $L$, the shift cannot be larger than $ \delta_{\max} \tilde{I}$, such that $ \tilde{I}(\omega \to \infty )+\delta_{\max} \tilde{I}=1$.
As a consequence, the integrated spectral function in the thermodynamic limit is bounded from below by a finite-size $ \tilde{I}(\omega)$, and from above by $\tilde{I}(\omega) + \delta_{\max} \tilde{I}$.
At weak disorder, $\delta_{\max} \tilde{I}$ is sufficiently small so that that the latter bound provides a reasonable estimate of $ \tilde{I}(\omega)$ in an infinite system.
The region within the bounds is marked in Fig.~\ref{figs2}(b) and~\ref{figs2}(d)  as a shaded area.

\begin{table}[]
    \centering
    \begin{tabular}{c|c|c}
    \ \   \ \ W   \ \  \ \  &  \ \   \ \  L   \ \   \ \ &  \ \   \ \ $\delta \tilde{I}$  \ \   \ \  \\
    \hline
    1.5 & 12  & 0.036 \\
    1.5 & 14  & 0.014 \\
    2 & 12  & 0.059 \\
    2 & 14  & 0.027 \\
    2.5 & 12  & 0.056 \\
    2.5 & 14  & 0.028 \\
    4 & 12  & 0.012 \\
    4 & 14  & 0.0055 \\
    \end{tabular}
    \caption{The vertical shifts of $\tilde{I}$ used in the inset in Fig.~\ref{fig1}(b) [main text] and in Figs.~\ref{figs2}(b),~\ref{figs2}(d) and~\ref{figs2}(f).}
    \label{tabS1}
\end{table}

\section{The role of the Lorentzian broadening} \label{app2}

In the main text we argued that the spectral function $S_{M}(\omega)$ from Eq.~(\ref{mazur2}), in the regime
$\tau^{-1}_{\rm max}  \ll  \omega  \ll \tau^{-1}_{\rm min}$,
roughly scales as $S(\omega)\propto 1/\omega^\eta$, where $\eta$ is related to the exponent $\mu$ that characterizes the power-law distribution of the relaxation times: $\eta \simeq 2-\mu$ (at $\mu<2$).
Here we show that such relation is not necessary a consequence of the Lorentzian broadening used in the derivation of Eq.~(\ref{mazur2}), but may also occur when the Lorentzians are replaced by other delta sequences. 
For the simplest choice  
\mbox{$\delta(\omega) \to \tau_{\alpha} \theta(1-|\omega \tau_{\alpha} |)/2$} one may easily calculate  $S_{M}(\omega )$.
In this case, Eq.~(\ref{mazur2}) should be replaced with 
\begin{eqnarray}
S_{M}(\omega ) &=&  \frac{\bar D_0}{2}  \int_{\tau_{\rm min}}^{\tau_{\rm max}} \frac{{\rm d} {\tau}}{\tau^{\mu-1}}  \theta(1-|\omega \tau |) \nonumber \\
&=&  \frac{\bar D_0}{2}  \int_{\tau_{\rm min}}^{\omega^{-1}} \frac{{\rm d} {\tau}}{\tau^{\mu-1}} \nonumber \\
&=& \frac{\bar D_0}{2(2-\mu)}\left(  \frac{1}{\omega^{2-\mu}}-\tau^{2-\mu}_{\rm min} \right) \nonumber \\
& \simeq & \frac{\bar D_0}{2(2-\mu)}\ \frac{1}{\omega^{2-\mu}}\;,
\label{mazur2s}
\end{eqnarray}
and the latter approximation holds true for  $\omega  \ll \tau^{-1}_{\rm min}$. 
One observes that the Lorentzian-broadening (with broad high-frequency tails) and the rectangular-broadening (where the high-frequency part is absent) lead to the same frequency dependence of the spectral function,
$S_{M}(\omega) \propto \frac{1}{\omega^{2-\mu}}$.
It demonstrates that the details of the delta-function broadening are not essential for $S_{M}(\omega )$.

Nevertheless, several numerical studies have recently observed a Lorentzian form of the spectral function in models close to integrable points~\cite{mierzejewski2015, schoenle_jansen_21, leblond2020}.
Moreover, the Lorentzian form of the spectral function for spin imbalance is consistent with the standard diffusion~\cite{prelovsek2021}.
We consider a system using fermionic representation, which at time $t=0$ has spatially periodic distribution of particles, $n_i(0)=C_0 \cos(q i)$.
In the diffusive regime, the amplitude decays exponentially in time,
$C(t)=C_0 \exp[-D_q q^2 t]$,
where the diffusion constant is $D_{\rm diff} = \lim_{q \to 0} D_q$.
Then, the Fourier transform of $C(t)$ is a Lorentzian. The same is expected also for the spin imbalance studied here, which in the fermionic representation reads
$\hat A \propto \sum_i \cos(\pi i) (\hat{n}_i-1/2) \propto \hat{n}_{q=\pi}$.  

\section{Details about the fitting} \label{app3}

\begin{figure}[!b]
\centering
\includegraphics[width=\columnwidth]{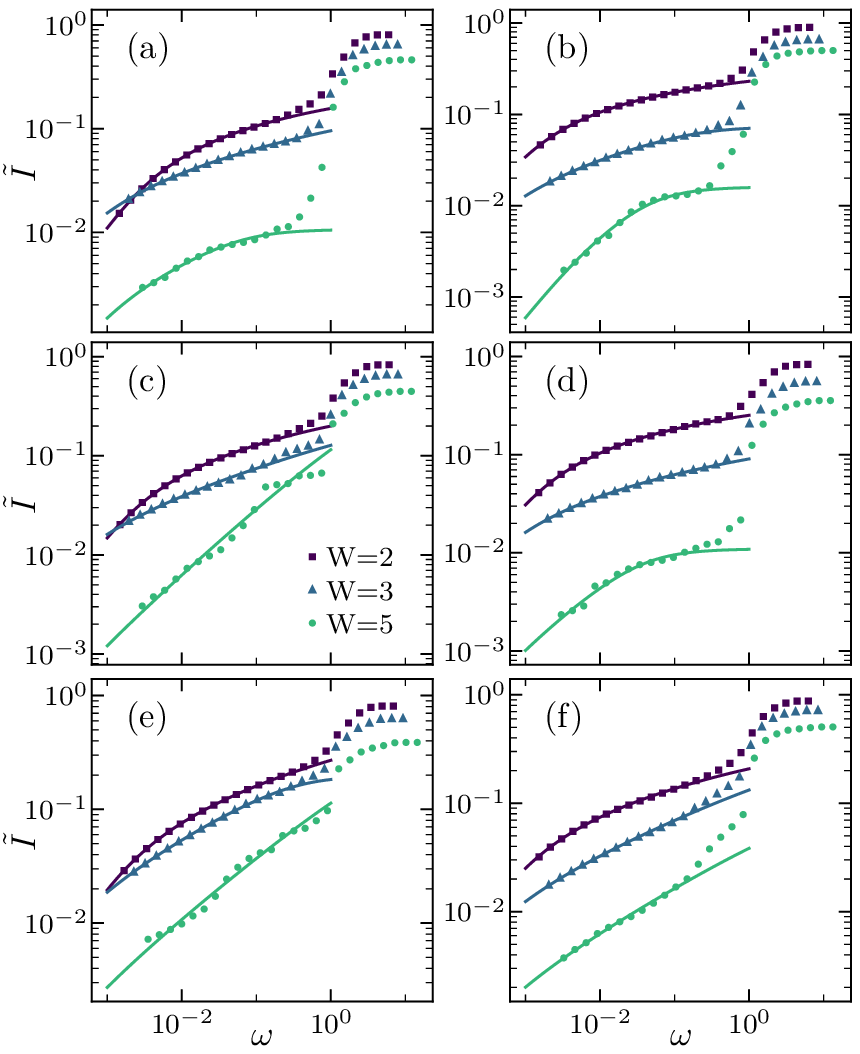}
\caption{
Symbols: numerical results for $\tilde{I}(\omega)$ at $L=16$ and a single realization of the disorder.
Lines: predictions by $\tilde{I}_M(\omega)$ for the low-frequency regime $\omega<0.2$.
Different panels correspond to different realizations of the disorder.
} 
\label{figs3}
\end{figure} 

So far most of the analytical considerations focused on properties of the spectral function $S_{M}(\omega)$ from Eq.~(\ref{mazur2}).
The function that we actually fit to the numerical values of $\tilde I(\omega)$ is
\begin{eqnarray}
\tilde{I}_M(\omega) & = & \int_{-\omega}^{\omega} {\rm d} \omega'  \frac{\bar D_0 }{\pi}  \int_{\tau_{\rm min}}^{\tau_{\rm max}} \frac{{\rm d} {\tau}}{\tau^{\mu-1}}  \frac{1}{(\omega' \tau)^2+1} \nonumber \\
&=& \frac{2 \bar D_0 }{\pi} 
\int_{\tau_{\rm min}}^{\tau_{\rm max}} 
\frac{{\rm d} {\tau}}{\tau^{\mu}}  {\rm arctan}(\omega \tau)\;,
\label{itomegas}
\end{eqnarray}
where the fitting parameters are $\mu$, $\tau_{\rm min}$ and $\tau_{\rm max}$ that determine the distribution of relaxation times $f_\tau(\tau)$, and the prefactor $\bar D_0$.
Since the results span over a few orders of magnitude, we fit
${\rm log}[ \tilde{I}_M(\omega)]$ to ${\rm log}[ \tilde{I}(\omega)]$ for $\omega <0.2$.
We bound the parameters $0<\mu<2$, $0.1<\tau_{\rm min}<20$, $20<\tau_{\rm max}<\tau_{\rm max}^{(\infty)}$ and $0 < \bar D_0 < 1$.
At $L=16$, $\tau_{\rm max}^{(\infty)}$ is either $10^4$ or infinity (see the discussion below).
In the case when $\tau_{\rm max} \to \infty$, the exponent $\mu$ is also bounded from below, $\mu>1$, otherwise $f_\tau(\tau)$
can not be properly normalized.
For smaller systems $L=14$ and $L=12$ this
bound is rescaled, respectively, down to $2666$ and $718$, 
so that $\tau_{\rm max}^{(\infty)}/{\cal D}$ is
the same for all system sizes.

\begin{figure}[!]
\centering
\includegraphics[width=\columnwidth]{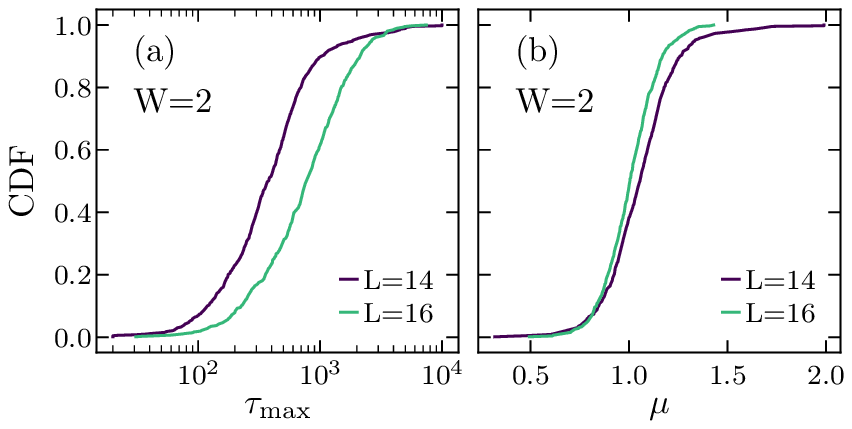}
\caption{
The cumulative distribution functions (CDF) of the fitting parameters $\tau_{\rm max}$ in (a) and $\mu$ in (b).
The fitting is performed independently for each realization of the disorder.
We perform the fitting for $10^3$ realizations of the disorder at $W=2$ and $L=14, 16$.
} 
\label{figs4}
\end{figure} 

\subsection{Fitting results for a single disorder realization} \label{app3a}

The main advantage of studying the {\it integrated} spectral function $\tilde{I}(\omega)$ is that one may analyze results obtained for various realizations of the disorder without averaging over them.
The fits of the phenomenological model $\tilde I_M(\omega)$ [lines] to numerical results $\tilde I(\omega)$ [symbols] is shown for a single realization in Fig.~\ref{fig3}(a) in the main text, and for six other realizations in Fig.~\ref{figs3}.
In all the cases, the agreement is excellent.

We carried out, in total, the fitting procedure for $10^3$ realizations of the disorder and studied the distributions of the fitting coefficients $\tau_{\rm max}$ and $\mu$.
First, we note that the distributions of $\tau_{\rm max}$ are very broad [cf.~Fig.~\ref{fig3}(b) of the main text and Fig.~\ref{figs4}(a)], i.e., the realization-to-realization fluctuations may differ by an order of magnitude.
Second, we observe that the average of $\tau_{\rm max}$ increases when both $W$ or $L$ are increased.
The increase with $W$ is shown in Fig.~\ref{fig3}(b) of the main text, while the increase with $L$ at $W=2$ is shown in Fig.~\ref{figs4}(a).
This dependence is discussed in more detail below.
We note that at $W = 2$ (i.e., when $\tau_{\rm max} \approx t_{H}=\omega^{-1}_{H}$), the distribution of $\mu$ is peaked around $\mu = 1$, and it exhibits only a weak $L$ dependence, see Fig.~\ref{figs4}(b).

\subsection{Fitting results for disorder averages} \label{app3b}

\begin{figure}[!]
\centering
\includegraphics[width=\columnwidth]{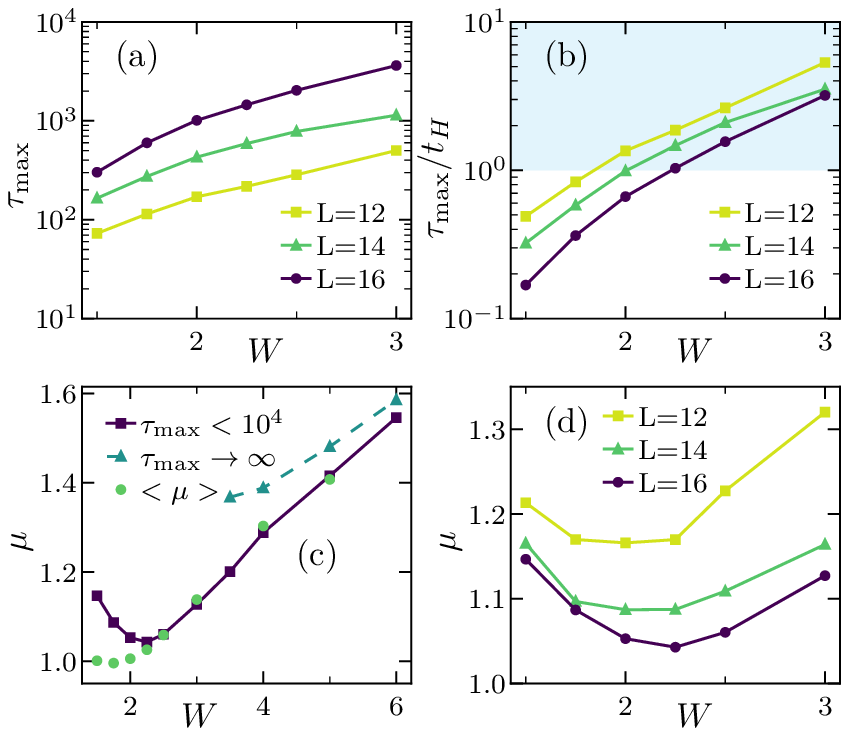}
\caption{
Results from spectral functions which were averaged over $10^3$ realizations of disorder.
(a) and (b) $\tau_{\rm max}$ and $\tau_{\rm max}/t_{H}$, respectively, as a function of the disorder $W$ at $L=12, 14, 16$.
The shaded region in (b) marks the regime $\tau_{\rm max}/t_{H} > 1$.
Unless stated otherwise, we obtain $\tau_{\rm max}$ and $\mu$ by fitting the function $\tilde I_{M}(\omega)$ from Eq.~(\ref{itomegas}) to the disorder averaged numerical values of $\tilde{I}(\omega)$, as explained in Sec.~\ref{app3b}.
(c) $\mu$ vs $W$ at $L=16$.
The upper bound for $\tau_{\rm max}$ is either $\tau_{\rm max}^{(\infty)} = 10^4$ (squares) or $\tau_{\rm max}^{(\infty)} = \infty$ (triangles).
Circles correspond to the averages of distributions of $\mu$ obtained from the fitting procedure described in Sec.~\ref{app3a}.
(d) $\mu$ vs $W$ at $L=12, 14, 16$.
} 
\label{figs5}
\end{figure} 

We complement previous results by studying the results of fitting the function $\tilde I_{M}(\omega)$ from Eq.~(\ref{itomegas}) to the disorder averaged numerical values of $\tilde{I}(\omega)$.
The latter are averaged over $10^3$ realizations of the disorder.
We obtain an excellent agreement between $\tilde I_{M}(\omega)$ and $\tilde I(\omega)$, as shown in Fig.~\ref{fig4} in the main text.
Here we comment on the values of the fitting parameters $\tau_{\rm max}$ and $\mu$.

We observe several interesting features of $\tau_{\rm max}$ (we focus on $L=16$).
It increases very rapidly (approximately exponentially) with $W$ and it reaches the Heisenberg time $t_{H}$ at $W^* \approx 2$, see Figs.~\ref{figs5}(a) and~\ref{figs5}(b).
When $\tau_{\rm max} > t_{H}$, the diffusive character of the dynamics in a finite system disappears completely.
One may argue that $\tau_{\rm max}$ quantitatively resembles the scaling of the Thouless time $t_{\rm Th}$ obtained from the spectral form factor~\cite{suntajs_bonca_20a}.
Intriguingly, the criterion $t_{\rm Th} \approx t_{H}$ provides an accurate tool to pinpoint the Anderson localization transition in three dimensions studied by the spectral form factor~\cite{suntajs_prosen_21, sierant_delande_20}.
Despite this similarity, we note that $\tau_{\rm max}$ was introduced as a fitting parameter of the phenomenological model in Eqs.~(\ref{mazur2}) and~(\ref{itomegas}), with no apparent formal similarity with $t_{\rm Th}$.
It is important to stress that in the regime $\tau_{\rm max} > t_{H}$, the quality of the fits does not strongly depend on $\tau_{\rm max}$.
This uncertainty of $\tau_{\rm max}$  is marked by the shaded area in Fig.~\ref{figs5}(b).

The relevance of the above discussion can also be seen in the analysis of $\mu$ in Fig.~\ref{figs5}(c).
There are two lines in Fig.~\ref{figs5}(c) at $W>3$: the dashed line (with triangles) corresponds to the results for $\mu$ when 
 $\tau_{\rm max}$ is sent to infinity (i.e, $\tau_{\rm max}$ is not a fitting parameter any more), while the solid line (with squares) corresponds to the results for $\mu$ when $\tau_{\rm max}^{(\infty)}=10^4$ (as an estimate, $t_{\rm H} \approx 10^3$ at $L=16$).
A reasonable agreement between both lines confirms that the choice of $\tau_{\rm max}$ at large $W$ is less important, provided that it satisfies $\tau_{\rm max} > t_{\rm H}$.

The main goal of this work is to establish a phenomenological model to describe the low-frequency dynamics, based on the proximity to the Anderson insulator and the emergent power-law distribution of relaxation times of the Anderson LIOMs.
A quantitative determination of the power-law exponent $\mu$ of the relaxation time distribution in the thermodynamic limit is beyond the scope of this work.
Still, in Figs.~\ref{figs5}(c) and~\ref{figs5}(d) we report some properties of $\mu$ as a function of $W$ and $L$.
We first stress that in the regime $W < W^* \approx 2$, the bounds $\tau_{\rm min}$ and $\tau_{\rm max}$ of the distribution may still be quantitatively close to each other and hence the determination of $\mu$ is more ambiguous.
This can be seen in Figs.~\ref{figs5}(c) and \ref{figs5}(d) as the departure of $\mu$ from $\mu=1$ when fitting the results for the disorder averaged $\tilde I(\omega)$ [solid line with squares in Fig.~\ref{figs5}(c)].
In contrast, the mean value of $\mu$ obtained after fitting results for every disorder realization separately remains very close to 1 when $W < W^*$ [circles in Fig.~\ref{figs5}(c)].
In the opposite regime $W > W^*$, $\mu$ increases as a function of $W$ for both types of fitting procedure.
However, as argued above, in this regime the width of the power-law distribution of relaxation times is larger than the range of numerically accessible frequencies, and hence the flow of $\mu$ when approaching the thermodynamic limit may be ambiguous.
Finally, in Fig.~\ref{figs5}(d) we show results for $\mu$ in the vicinity of $W \approx W^*$ for the three system sizes $L=12,14,16$.
When increasing $L$ the value of $\mu$ shrinks to lower values, and it eventually approaches the regime $\mu \approx 1$, at least for the given interval of disorders.

\section{Anderson LIOMs in interacting systems at $\Delta > 0$} \label{appnew}

Our phenomenological approach that quantitatively describes the dynamics of the imbalance in the random field Heisenberg chain is based on an assumption
that (at least some) Anderson LIOMs, $\hat Q_{\alpha}$, decay in interacting systems ($\Delta > 0$) with a finite relaxation time $\tau_{\alpha }$,
and that  $\tau_{\alpha }$ are random variables with a broad, power-law distribution. Moreover,  the projections of the spin imbalance on various Anderson LIOMs [see Eq.~(\ref{mazur1}) in the main text] have been approximated  by the average projection. In this section, we present numerical results that directly support these conjectures and approximations.

For convenience we study the fermionic model,
\begin{eqnarray}
\hat H &=&  \hat H_0 + J \Delta  \sum_i    \hat{n}_i  \hat{n}_{i+1} , \label{hamhs} \\
\hat H_0 &=& \frac{J}{2} \sum_i  \left( \hat a^{\dagger}_{i+1} \hat a_{i}+{\rm H.c.} \right) +  \sum_i  h_i \hat{n}_i  , \label{hamh0s},
\end{eqnarray}  
which is, up to a constant term, equivalent to the Hamiltonian~(\ref{hamh}) in the main text. Here, $\hat a^{\dagger}_{i}$ creates a spinless fermion at site $i$ and  
$ \hat{n}_i = \hat a^{\dagger}_{i} \hat a_{i}$.

\begin{figure}[!]
\centering
\includegraphics[width=\columnwidth]{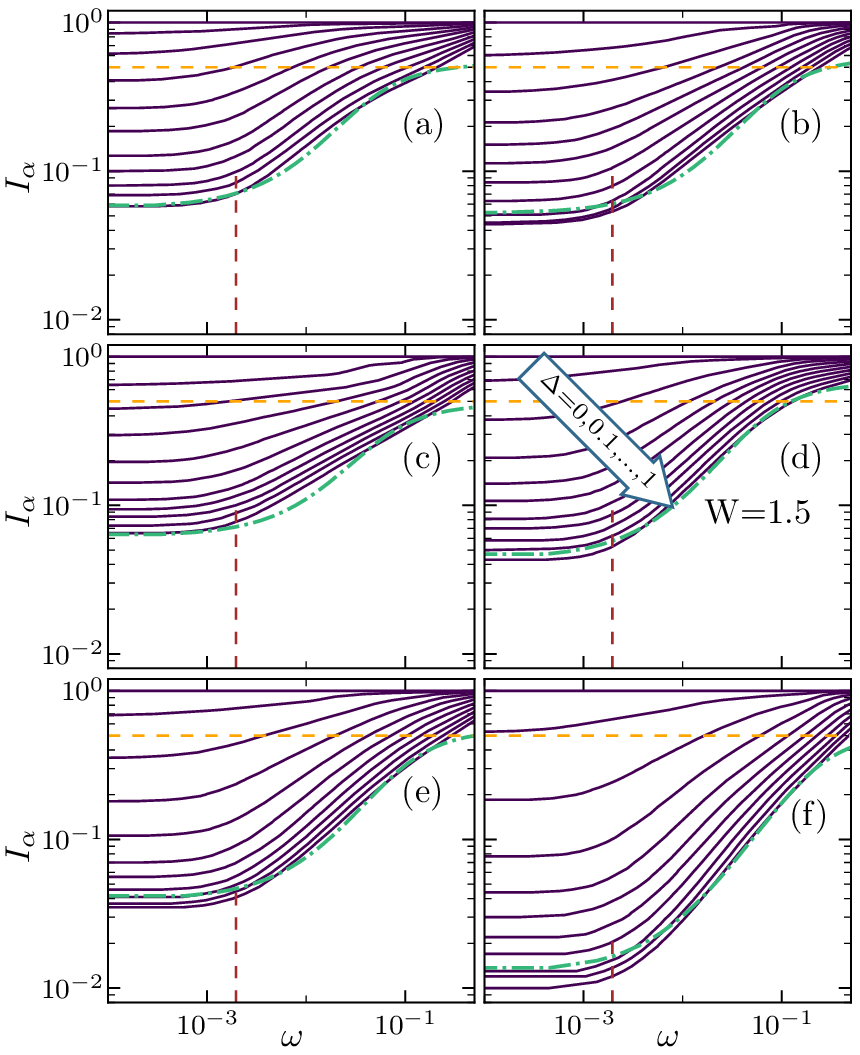}
\caption{Integrated spectral functions of the Anderson LIOMs, Eq.~(\ref{sitomega}), at
$W=1.5$ and $L=14$. Each panel shows results for one Anderson LIOM and a single realization of disorder. Various curves have been obtained for
$\Delta=0,0.1,...,1$, as it is indicated by an arrow in panel (d).
The frequency for which $I_{\alpha}(\omega)=1/2$, see the horizontal  dashed lines, defines the
relaxation rate $\Gamma_{\alpha}=\omega$ via Eq.~(\ref{staua}). Vertical dashed curves mark the Heisenberg energy, $\omega_{H}=1/t_{H}$, calculated at $\Delta=1$. Dashed-dotted (green)  lines
show a low-frequency fit ($\omega < 0.3$) to the numerical results at $\Delta=1$. Here, we have used the fitting function $I_{\rm fit}=f_1 \arctan(f_2\;\omega)+f_3$, where $f_i$ are the fitting parameters.
}
\label{figs6}
\end{figure} 

For each configuration of the disorder, we determine the Anderson states $| \alpha \rangle $ and the relevant operators, $\hat a^{\dagger}_{\alpha}=\sum_i \langle i | \alpha \rangle \hat a^{\dagger}_i $, which diagonalize the single particle Hamiltonian~(\ref{hamh0s}), $\hat H_0=\sum_{\alpha}  \epsilon_{\alpha} \hat a^{\dagger}_{\alpha} \hat a_{\alpha}$.
Then, using the full Hamiltonian from Eq.~(\ref{hamhs}) we study the  dynamics of the one-body Anderson LIOMs, $\hat Q_{\alpha}=2 (\hat a^{\dagger}_{\alpha} \hat a_{\alpha}-\frac{1}{2})$, which are normalized and mutually orthogonal, $\langle  \hat Q_{\alpha} \hat Q_{\alpha'}\rangle=\delta_{\alpha, \alpha'} $.
Here, we do not consider the products of $\hat Q_{\alpha}$  (e.g.,  $\hat Q_{\alpha} \hat Q_{\alpha'} \hat Q_{\alpha''}$) even though they may also  contribute to the Mazur bound [Eq.~(\ref{mazur}) in the main text], especially at weak disorder.
In analogy to Eqs.~(\ref{iomega}) and~(\ref{itomega}) in the main text, for each realization of the disorder and each $\hat Q_{\alpha}$ we determine the spectral functions $S_{\alpha}(\omega)$ and the integrated spectral functions $I_{\alpha}(\omega)$,
\begin{eqnarray}
S_{\alpha}(\omega)&=&\frac{1}{2 \pi} \int_{-\infty}^{\infty} {\rm d} t \; e^{i \omega t -|t| 0^+} \langle e^{i\hat Ht} \hat Q_{\alpha}   e^{-i\hat Ht} \hat  Q_{\alpha} \rangle\,, \label{siomega} \\ 
I_{\alpha}(\omega)&=&\int_{-\omega}^{\omega} {\rm d} \omega'  S_{\alpha} (\omega') \nonumber \\
&=& \frac{1}{\cal D} \sum_{m,n=1}^{\cal D} \theta \left( \omega-|E_m-E_n| \right) \langle m | \hat Q_{\alpha} | n \rangle ^2 \,. 
\label{sitomega}
\end{eqnarray} 
Figure \ref{figs6} shows $I_{\alpha}(\omega)$ where each panel contains results for a single $\alpha$ and one realization of disorder.
Various curves demonstrate how  the integrated spectral function changes upon increasing $\Delta$ starting from  the noninteracting system at $\Delta=0$.
In the latter case, $I_{\alpha}(\omega)$ is a step function since  $\hat Q_{\alpha}$ are strictly conserved.  However, $I_{\alpha}(\omega)$ broadens at $\Delta > 0$ reflecting the onset of a finite relaxation time $\tau_{\alpha}$. In the low-frequency regime, this broadening may be reasonably well fitted by  $I_{\rm fit}=f_1 \arctan(f_2\;\omega)+f_3$,  see the dashed-dotted lines, in accordance with the Lorentzian broadening introduced in
Eq. (\ref{mazur1}) in the main text.  Here, the fitting parameter $f_3$ reproduces the saturation of the spectral function when the frequency is smaller than the Heisenberg energy $\omega_{\rm H}$.

We quantitatively obtain the relaxation time $\tau_{\alpha}$ by approximating the autocorrelation function by an exponential function,
$\langle e^{i\hat Ht} \hat Q_{\alpha}   e^{-i\hat Ht} \hat  Q_{\alpha} \rangle \propto \exp(-t/\tau_{\alpha})$, which using Eqs.~(\ref{siomega}) and~(\ref{sitomega}) implies that 
\begin{equation}
I_{\alpha}\left(\omega=\frac{1}{\tau_{\alpha}} \right)=\frac{1}{2} \label{staua} \,.
\end{equation}
Solving Eq. (\ref{staua}) allows for a simple numerical extraction of the relaxation rate, $\Gamma_{\alpha}= \frac{1}{\tau_{\alpha}}$, for each realization of the disorder and each $\hat Q_{\alpha}$.

\begin{figure}[!]
\centering
\includegraphics[width=\columnwidth]{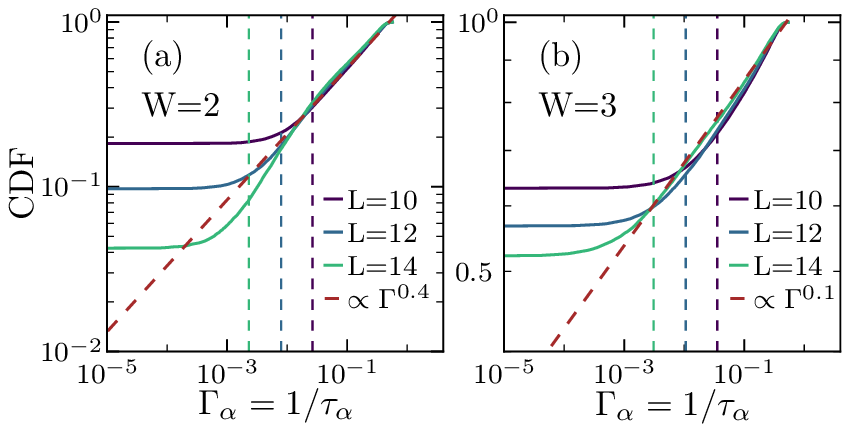}
\caption{
Cumulative distribution functions (CDF) of the relaxation rates, $\Gamma_{\alpha}=\frac{1}{\tau}_{\alpha}$, defined via Eq.~(\ref{staua}).
Results are  obtained for $10^3$ realizations of the disorder and for all one-body Anderson LIOMs $\hat Q_{\alpha}$, with $\alpha=1,...,L$. The dashed vertical lines mark the Heisenberg energy $\omega_{H}=1/t_{H}$.}
\label{figs7}
\end{figure} 

Figures~\ref{figs7}(a) and~\ref{figs7}(b) show the cumulative distribution functions (CDF) of the relaxation rates, obtained at $W=2$ and $W=3$, respectively.
The distributions have been obtained from $10^3$ realizations of the disorder and for all one-body Anderson LIOMs, $\alpha=1,...,L$.
The verticals lines mark the values of the inverse Heisenberg time $\omega_{\rm H}$.
One observes that the relaxation rates obtained for various realizations of the disorder may differ by a few orders of magnitude.
Results in Fig.~\ref{figs7}
allow us also to test the conjecture that the probability density for the relaxation times is $f_{\tau}(\tau) \propto 1/\tau^\mu$ with $\mu < 2$. The CDF of $\Gamma_{\alpha}$ is related to $f_{\tau}(\tau)$ via the following relation 
\begin{eqnarray}
{\rm CDF}_{\Gamma}&=&\int_{\Gamma^{-1}}^{\tau_{\rm max}} {\rm d} \tau f_{\tau}(\tau) \; \propto \; \left( \Gamma^{\mu-1} -\frac{1}{\tau_{\max}^{\mu-1}} \right).
\end{eqnarray}
It means that for the assumed distribution, $f_{\tau}(\tau)$, one expects ${\rm CDF}_{\Gamma} \propto \Gamma^{\mu-1} $ for $\Gamma \gg \tau^{-1}_{\max}$.
Figures~\ref{figs7}(a) and~\ref{figs7}(b) show that 
at $\Gamma > \omega_{H}$ we indeed observe the power-law form of ${\rm CDF}_{\Gamma}$  with $\mu \simeq 1.4$ and  $\mu \simeq 1.1$ at $W=2$ and $W=3$, respectively. The latter values of the exponent $\mu$ reasonably agree with results in Figs.~\ref{figs5}(c) and~\ref{figs5}(d) in the preceding section.
 
 \begin{figure}[!]
\centering
\includegraphics[width=\columnwidth]{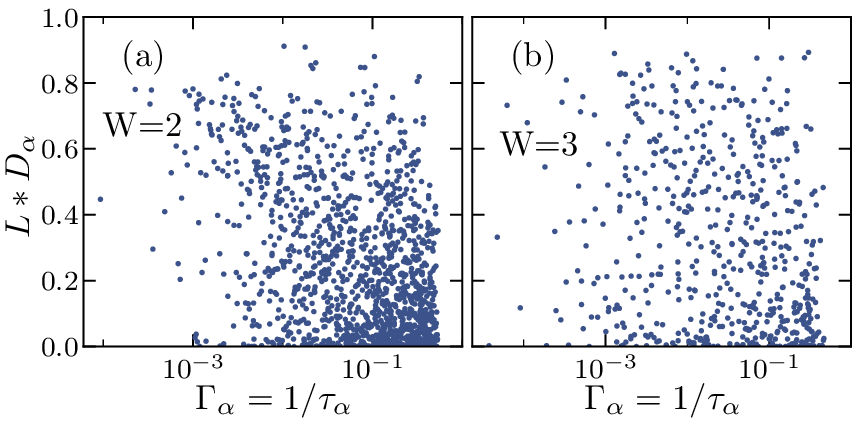}
\caption{
Correlations between the projections of spin imbalance on the Anderson LIOMs, $D_{\alpha}$,
[see Eq.~(\ref{mazur}) in the main text] and the relaxation rates $\tau_\alpha$ of the Anderson LIOMs [see Eq.~(\ref{staua}))]. Various points correspond to different realizations of the disorder or different Anderson LIOMs. Results have been obtained at $L=14$ and (a) $W=2$ and (b) $W=3$.}
\label{figs8}
\end{figure} 
 
In the main text we have also assumed that the projections of the spin imbalance on the Anderson LIOMs [$D_{\alpha}$ in Eq.~(\ref{mazur1}) in the main text] are not essential and can be replaced by an average value $D_{\alpha} \simeq $ const. The minimal requirement for this approximation to hold true is the absence of any significant correlations between $\tau_{\alpha}$  and $D_{\alpha}$.
Figures \ref{figs8}(a) and \ref{figs8}(b)  show the pairs of both quantities ($\Gamma_{\alpha}, D_{\alpha}$)  obtained for various realizations of the disorder and various $\alpha$. For a broad range of the relaxation rates, $10^{-4}<\Gamma_{\alpha} <10^{-1}$, the projections seem to cover the entire window of accessible values of $D_{\alpha}\in (0,L^{-1})$. Therefore, we expect that the approximation that decouples the relaxation times from  $D_{\alpha} $  does not introduce any significant error to the dynamics of the spin imbalance.


\end{document}